\documentclass[a4paper]{elsarticle}

\usepackage{lineno,hyperref,aas_macros}
\modulolinenumbers[5]
\def\astrobj#1{#1}
\journal{Journal of \LaTeX\ Templates}

\begin{document}

\begin{frontmatter}

\title{Asteroseismology of the ultramassive ZZ Ceti star \astrobj{WD~0246+326}}
\tnotetext[mytitlenote]{Based on data obtained
at the Xinglong Station of National Astronomical Observatories of China and San P\'{e}dro Mart\'{i}r Observatory of Mexico.}

\author[mymainaddress]{Chun Li}


\author[mymainaddress]{Jianning Fu\corref{mycorrespondingauthor}}
\cortext[mycorrespondingauthor]{Corresponding author}
\ead{jnfu@bnu.edu.cn}

\author[mysecondaryaddress]{Lester Fox-Machado}
\author[mymainaddress]{Jie Su}
\author[mymainaddress]{Fangfang Chen}

\address[mymainaddress]{Department of Astronomy, Beijing Normal University, Beijing, China}
\address[mysecondaryaddress]{Observetorio Astr\'{o}mico Nacional, Instituto de Astronomia, Universided Nacional Aut\'{o}noma de M\'{e}xico, Ensenada, M\'{e}xico}

\begin{abstract}
The internal structures of pulsating white dwarfs can be explored only with asteroseismology. Time series photometric observations were made for the pulsating DA white dwarf (ZZ Ceti star) \astrobj{WD~0246+326} during 9 nights in 2014 with a bi-site observation campaign. Eleven frequencies were detected including 1 triplet, 2 doublets, and 4 single modes, which are identified as either $l=1$ or $l=2$ modes with the complementarity of frequencies present in the literature. From the multiplets, the rotation period of \astrobj{WD~0246+326} is derived as $3.78\pm 0.11$ days. The average period spacing of the l=1 modes $\Delta P=29.3\pm 0.2s$, implies that \astrobj{WD~0246+326} may be a massive ZZ Ceti star concerning the $\Delta P-M_*$ relationship for the DAVs. Preliminary analysis derives the stellar parameters of $M_*=0.98\pm0.01$~${\rm M_\odot}$ and $T_{\rm eff}=11700\pm100$~K by fitting the theoretical frequencies of the eigen modes to the observed ones.

\end{abstract}


\end{frontmatter}

\section{Introduction}
As final remains of stellar evolutions of the majority of stars in the Galaxy, white dwarfs offer opportunities of investigations with significant constraints on ages of both the Galactic Disk and the globular clusters (Winget et al. 1987, 2008\cite{1987ApJ...315L..77W,2008ARA&A..46..157W}). They are also nature laboratories for testing physics under extreme conditions. Landolt(1968)\cite{1968ApJ...153..151L} discovered the first pulsating white dwarf \astrobj{HL~Tau~76}. As asteroseismology is a unique and powerful method known to explore the internal structures of stars, the pulsating white dwarfs have been hot targets of asteroseismology particularly due to their relatively simple internal structures compared to the stars at the other phases of evolution. Pulsating DA white dwarfs, also called ZZ Ceti stars, have the lowest temperatures among all four kinds of pulsating white dwarfs (Saio 2013\cite{2013EPJWC..4305005S}). The instability strip of ZZ Ceti stars locates at the intersection of the Cepheid instability strip and the evolution track of white dwarfs. The purity of the instability strip makes ZZ Ceti stars excellent samples to investigate the internal structures of all DA white dwarfs. However, although they compose the biggest group of pulsating white dwarfs with the largest number of confirmed members by far, the properties of low luminosities, low amplitudes and high frequencies enhance the difficulties of making observational investigations on these stars.

Statistical studies show that ZZ Ceti stars with low effective temperatures locate close to the red edge of the instability strip with long periods, high amplitudes and large amplitude modulations, while the ones with high effective temperatures exhibit the opposite properties. Short time scale variations of pulsating parameters (mainly the amplitudes) are reported for multiple ZZ Ceti stars through time series observations (c.f. Bogn\'{a}r et al. 2009\cite{2009MNRAS.399.1954B}, Provencal et al. 2012\cite{2012ApJ...751...91P}, Fu et al. 2013\cite{2013MNRAS.429.1585F}).

The variability of the ZZ Ceti star \astrobj{WD~0246+326} (\astrobj{KUV~02464+3239}) was discovered by Fontaine, Bergeron \& Bill\`{e}res (2001)\cite{2001ApJ...557..792F}, who reported only one pulsating mode with a period of 832~s from the Fourier transform due to the short duration of data ($\sim$ one hour of observations). Two sets of stellar parameters were given by Bergeron et al. (1995)\cite{1995ApJ...449..258B} and Gianninas et al. (2011)\cite{2011ApJ...743..138G} respectively, including the effective temperatures T$_{\rm eff}$ and the surface gravities $Log$ $\rm{g}$.

Bogn\'{a}r et al. (2009)\cite{2009MNRAS.399.1954B} carried out photometric observations from Konkoly Observatory with a 1-meter telescope in 2006 and 2007. Light curves of \astrobj{WD~0246+326} during $\sim$ 97 hours were obtained and six modes were reported.

We made bi-site time-series photometric observations for \astrobj{WD~0246+326} in 2014. The descriptions of the observations and data reduction is introduced in Section 2. Section 3 presents the process of frequency analysis. In Section 4 we focus on asteroseismological analysis. Our modeling attempt and related results are provided in Section 5. Finally, we make discussion and give conclusions in Section 6.
\section{Observations and Data Reduction}
 A bi-site observation campaign was carried out in October of 2014. Table~1 lists the journal of observations. The 2.16-m telescope at Xinglong (XL) Station of National Astronomical Observatory of China and the 1.5~m reflector of San P\'{e}dro Mart\'{i}r Observatory (SPM) of Mexico were used during the campaign. The instrument BFOSC mounted to the 2.16-m telescope of XL was taken as the detector through the Johnson $V$ filter. The data at SPM were acquired with the instrument RATIR (Reionization and Transients Infrared Camera, Butler et al. 2008\cite{2012SPIE.8446E..10B}), which consists of two optical and two infrared channels, simultaneous imaging in six colors, two optical and four near-infrared (http://ratir.astroscu.unam.mx/public/). Only the two optical cameras were used in this campaign. The frames were obtained simultaneously through the Bessel $V$ and $I$ filters (only frames observed though V filter were employed in further study). The number of images and data lengths during each observation night are listed in the 3rd and 4th columns of Table 1, respectively. The package of IRAF DAOPHOT was used to reduce the data according to the standard data reduction procedure. Light curves of \astrobj{WD~0246+326} are obtained and shown in Figure 1.

\begin{table}
 \centering
  \caption{Journal of observations in $V$ for \astrobj{WD~0246+326} in October of 2014. XL = Xinglong station of National Astronomical Observatories of China. SPM = San P\'{e}dro Mart\'{i}r Observatory of Mexico.}
  \begin{tabular}{@{}cccc@{}}
    \hline\noalign{\smallskip}
 Observatory/ & Date  & Frame  & Length \\
Telescope &&Number&(hour)\\
\hline &24& 280& 6.4\\
  &25& 468& 9.5\\
XL/ &26& 397& 9.2\\
2.16m &27& 436& 9.9\\
  &28& 83& 1.2\\
\hline &28& 236& 3.1\\
SPM/ &29& 231& 3.0\\
1.5m &30& 230& 2.9\\
  &31& 210& 2.7\\
\hline
\end{tabular}
\end{table}

\begin{figure}
\resizebox{\hsize}{!}{\includegraphics[angle=90]{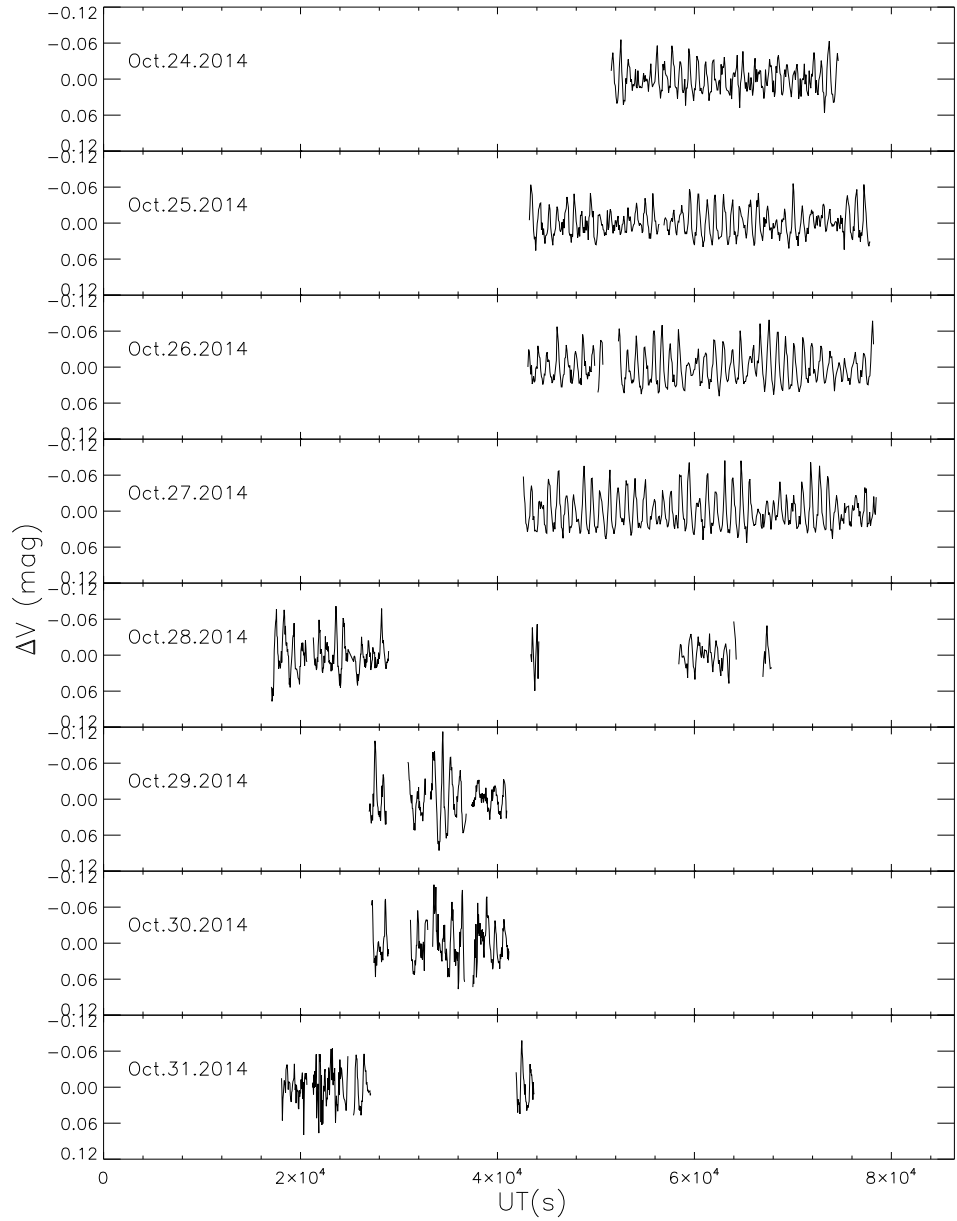}}
\caption{Light curves of \astrobj{WD~0246+326} in V in 2014.}
\label{fig1}
\end{figure}

\section{Frequency analysis}
\subsection{Frequencies Extracted From Light Curves}
We used the software {\tt Period04} (Lenz \& Breger 2005\cite{2005CoAst.146...53L}) to analyze the light curves. Eleven frequencies with signal-to-noise ratios larger than 4.0 (Breger et al. 1993\cite{1993A&A...271..482B}, Kuschnig et al. 1997\cite{1997A&A...328..544K}) were extracted with the pre-whitening procedure and are listed in Table 2. No linear combination or aliasing frequency were detected. Figure 2 shows the Fourier transforms of the light curves of \astrobj{WD~0246+326} and the residuals after 11 frequencies are extracted. We used the method of Monte-Carlo simulation to estimate the uncertainties of the frequencies and amplitudes. For more details about the Monte-Carlo simulation, we refer to Fu et al. (2013).\cite{2013MNRAS.429.1585F}

\begin{figure}
\resizebox{\hsize}{!}{\includegraphics[angle=90]{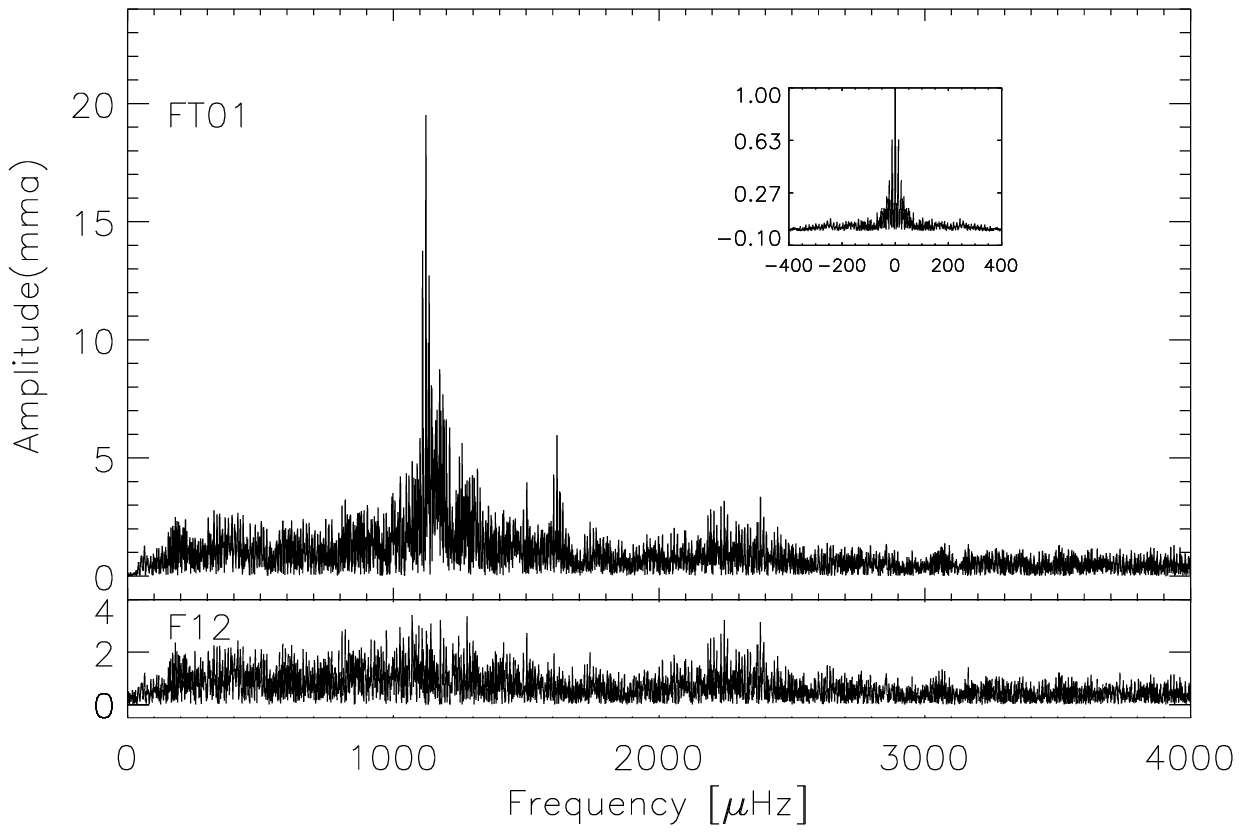}}
\vspace{4pt}
\caption{Fourier transforms of the light curves of \astrobj{WD~0246+326}. The upper panel shows the amplitude spectrum of the light curves with the spectral window shown as the insert. The lower panel shows the amplitude spectrum of the residuals with the 11 frequencies extracted through the prewhitening process.}
\label{fig2}
\end{figure}
\begin{table}
\caption{Frequency solution resolved from the light curves. Note that $f$ is the frequency in $\mu {\rm Hz}$, $A$ the amplitude in mma, S/N the signal-to noise ratio. ID is the name of the frequency}
\centering
\begin{tabular}{@{}cccc}
\hline  ID&$f\pm3\sigma(\mu {\rm Hz})$ &  $A\pm3\sigma$(mma)&S/N\\
\hline
a7   &   $1049.27\pm0.22$&   $5.1\pm1.2$&5.9 \\
a1   &   $1122.67\pm0.07$ &  $20.6\pm1.3$&24.2\\
a3   &   $1131.46\pm0.20$&   $5.8\pm1.6$&6.8\\
a2   &   $1174.21\pm0.22$&   $7.6\pm1.4$&9.0\\
a10  &   $1179.43\pm0.27$&   $4.8\pm1.4$&5.7\\
a11  &   $1181.93\pm0.32$ &   $4.3\pm1.4$&5.1\\
a5   &   $1211.75\pm0.27$&   $5.0\pm1.3$&6.0\\
a9   &   $1213.53\pm0.54$&   $3.5\pm1.4$&4.2\\
a6   &   $1259.24\pm0.16$&   $6.5\pm1.3$&7.8\\
a8   &   $1304.32\pm0.24$&   $5.0\pm1.3$&6.1\\
a4   &   $1616.03\pm0.19$&   $6.6\pm1.1$&8.1\\
\hline
\end{tabular}
\end{table}

\subsection{Frequency Analysis and Mode Identification}
Table 3 lists the 11 frequencies in Table 2 and the 6 frequencies reported in Bogn\'{a}r et al. (2009)\cite{2009MNRAS.399.1954B}. The frequency separations between the adjacent frequencies, the corresponding periods, and the period separations are also given in Table 3. The frequencies are separated into different groups for the convenience of discussion.
\begin{table}
\caption{Frequency list of \astrobj{WD~0246+326}. $f$ is the frequency in $\mu {\rm Hz}$, $\delta f$ the difference between the adjacent frequencies. $P$ the corresponding period in \emph{second}
 and $\Delta P$ the period differences. Note that the frequencies in "b" are taken from Bogn\'{a}r et al. No. is the Number of frequency group, ID the name of the frequency. (2009)\cite{2009MNRAS.399.1954B}.}
 \centering
\small
\begin{tabular}{@{}cccccc}

\hline  No.&ID&$f(\mu Hz)$ & $\delta f(\mu {\rm Hz})$ &$P$(s) & $\Delta P(s)$\\
\hline
 &  b1& \textit{799.84}  &  &     \textit{1250.3}  &\\
& & &206.97&&257.0\\
G1&b2&   \textit{1006.80}  &  & \textit{993.2}   &\\
 & & &42.47&&40.2\\
&a7   &   $1049.27$&  & 953.0&\\
&  & &73.40&&62.3\\
\hline & a1   &   $1122.67$ &  & 890.7& \\
 G2& & &8.80&&6.9 \\
 &a3   &   $1131.46$&    &883.8&\\
\hline & & &23.07&&17.7\\
  G3 & b3&    \textit{1154.53}  &        & \textit{866.2}   &\\
    &  & &19.68&&14.5\\
\hline & a2   &   $1174.21$&  &851.6& \\
  & & &5.22&&3.8 \\
G4 &a10  &   $1179.43$&   & 847.9& \\
 & & &2.50&&1.8 \\
&a11  &   $1181.93$ &  & 846.2& \\
\hline && &24.80&&17.4\\
    G5& b4&   \textit{1206.73}  &        & \textit{828.7}    &\\
 & & &5.02&&3.43\\
\hline &a5   &   $1211.75$&  & 825.3& \\
 G6& & &1.78&&1.2 \\
&a9   &   $1213.53$&   & 824.0& \\
\hline & & &45.72&&29.9\\
 &a6   &   $1259.24$&   & 794.1&\\
  &   & &26.75&&16.5\\
 G7 & b5&    \textit{1285.99}  && \textit{777.6}&  \\
 && &18.33&&10.9\\
&a8   &   $1304.32$&  & 766.7&\\
 & & &310.45&&147.4\\
\hline &   b6&  \textit{1614.77}  & & \textit{619.3}&  \\
 G8& & &1.27&&0.5\\
&a4   &   $1616.03$&   & 618.8&\\
\hline
\end{tabular}
\end{table}

\begin{table}
\caption{Mode identifications and further signals. $f$ is the frequency in $\mu {\rm Hz}$, $\delta f$ the frequency separation between the consecutive frequencies in $\mu {\rm Hz}$, and $P$ the period in second. The frequencies with IDs given in "b" are from Table 4 of Bogn\'{a}r et al. (2009)\cite{2009MNRAS.399.1954B}}
\centering
\begin{tabular}{@{}cccccccc}
\hline  ID&$f(\mu {\rm Hz})$ & $\delta f$ & $P(s)$ & $\Delta P(s)$&$\delta k$&$\delta m$  \\
\hline \multicolumn{7}{c}{$l=1$}\\
\hline

 a5&  1211.75    &     &  825.3&       &\\
                    &     & 1.78                    &&          \\
  a9&  1213.53  &&824.0&               &+7\\

&&&&29.9&&\\
  a6&  1259.24      &&  794.1&      &+6\\

&&&&27.4&&\\

  a8&  1304.32    &&  766.7&            &+5\\
&&&&147.4&&\\
  b6&  \textit{1614.77}  &&  \textit{619.3}&             &0\\
                    &         & 1.27                    &&          \\
 a4&   1616.03  &&  618.8&            &\\

\hline \multicolumn{7}{c}{$l=2$}\\
\hline
 a1&   1122.67     && 890.7&     &      &+2? \\
                  &     &   8.80              &       &  &+2&        \\
 a3&   1131.46        &&  883.8&    & &-2 or -1\\
&&&&32.2&&\\
 a2&   1174.21      && 851.6&        &  &+2 or +1 \\
           &     & 5.22                        &     \\
  a10&  1179.43        &&  847.9&     & 0&0 or -1\\
            &             & 2.50                        & &    \\
 a11&   1181.93      && 846.1&         & &-1 or -2\\
    \hline \multicolumn{7}{c}{Further sigmals}\\
    \hline
   b1& \textit{799.84}  &        & \textit{1250.3}&    &&\\
     b2&   \textit{1006.80}  &        & \textit{993.2}&    &&\\
  a7&  1049.27    &        & 953.0&     &&\\
    b3&    \textit{1154.53}  &        & \textit{866.2}&    &&\\
     b4&   \textit{1206.73}  &        & \textit{828.7}&   &&\\
  b5&    \textit{1285.99}  &        & \textit{777.6}&    &&\\
\hline
\end{tabular}

\end{table}

\begin{enumerate}
\item\textbf{G6 and G8} Looking at Table 3, one finds, it can be seen from column 3 that there exist two groups of very closely-spaced frequencies, G6 and G8, with frequency separations of $1.78~\mu {\rm Hz}$ and $1.26~\mu {\rm Hz}$, respectively.  In addition, G2 and G4 have frequency separations equal to or smaller than $8.8~\mu {\rm Hz}$. The reason that we do not include b4 into G6 is given in (iv).
\item\textbf{G4} The frequency spacings in G4 are $5.2~\mu {\rm Hz}$ and $2.5~\mu {\rm Hz}$, respectively, with the spacing ratio of $\sim$2:1. From the spherical symmetry, we suppose that they could be frequencies of the same mode of $l=2$ with $m$ values of either +2, +0, -1 or +1, -1, -2.
\item\textbf{G4 and G8}  According to the equation
\[
\sigma_{k,l,m}=\sigma_{k,l}+m\times (1-C_{k,l}) \Omega
\]
Where $C_{k,l}=1/l(l+1)$ in the asymptotic regime (Brickhill 1975\cite{1975MNRAS.170..405B}), the rotation splits of the $l=2$ modes are $\sim1.67$ times of the ones of the $l=1$ modes. Since we identified G4 as $l=2$, the average frequency split of the $l=2$ modes is derived as $~2.5~\mu {\rm Hz}$. Thus, the frequency split of $l=1$ modes should be $\sim1.5~\mu {\rm Hz}$. Since the frequency separation of G2 is as large as $8.8~\mu {\rm Hz}$, we interpret them as rotation splits of a $l=2$ mode rather than a $l=1$ mode.

\item\textbf{G5} The frequency splits of G6 ($1.78~\mu Hz$) and G8 ($1.26~\mu Hz$) are close to the estimated frequency split of the l=1 modes. We thus identify them as $l=1$ modes. Since the frequency separation of b4-a5 of $5.02~\mu {\rm Hz}$ is much larger than the average frequency split of $l=1$ modes, we do not identify b4 as a frequency split of G6.
\item\textbf{Period spacing of $l=1$ modes} From the asymptotic theory (Unno et al. 1979\cite{1979nos..book.....U}, Tassoul 1980\cite{1980ApJS...43..469T}), periods of a pulsation mode can be estimated as
\[
P_{l,k}\simeq\frac{2\pi^2k}{\sqrt{l(l+1)}}(\int^R_0\frac{N}{r}dr)^{-1}
\]
Where $N$ is the Brunt-V\"{a}is\"{a}l\"{a} frequency, $R$ the stellar radius. The integral part of the right side of the equation is fixed in one ZZ Ceti star. This equation suggests a uniform period spacing for the modes with the same $l$ value and consecutive $k$ values.
We notice that the period spacings of a6-a9 as well as a6-a8 are 29.9~s and 27.4~s, respectively, which could be an evidence showing that a8 and a6 are also $l=1$ modes. A linear fit is made to a9, a6, a8 and b6 with the $k$ values of +7, +6, +5 and 0, respectively concerning the period spacing values between the frequencies. Figure 3 shows the linear fitting.
\item\textbf{Period spacing of $l=2$ modes} As deduced from the asymptotic theory, the average period spacing of the $l=1$ modes is $\sim\sqrt{3}$ times as large as that of the $l=2$ modes. The average period spacing $29.3\pm0.2~{\rm{s}}$ of $l=1$ modes is calculated with the four corresponding modes. Hence the average period spacing of the $l=2$ modes should be $\sim$16.9~s. We find that the period spacing between the deduced $m=0$ modes of the two multiplets a1-a3 and a2-a10-a11 is 39.3~s, not far from twice the average period spacing of the $l=2$ modes.

\item\textbf{The rest frequencies} The $l$ values of the rest other observed frequencies can not be identified directly. We list these frequencies as further signals.
\end{enumerate}

We summarize the mode identification result in Table 4.
\section{Asteroseismology}
\subsection{Frequency Splitting And Rotational Period}
From the two doublets which are identified as $l=1$ modes in Table 4, the average frequency splitting due to rotation is $1.52\pm0.13~\mu {\rm Hz}$. With the two sets of $l=2$ modes identified, we notice that the frequency separation of a10-a11 is approximately the half of that of a2-a10, suggesting the $m$ values of a2, a10, a11 as either +2, 0, -1 or +1, -1, -2, respectively. The average frequency splitting due to rotation for the $l=2$ modes are hence derived as $2.51\pm 0.01~\mu {\rm Hz}$. We do not use G2 to estimate the frequency splitting of the $l=2$ modes, since the $m$ values of a1 and a3 can not be assigned reliably.

With the average rotation splitting of the $l=1$ modes of $1.52\pm0.13~\mu {\rm Hz}$ and that of the $l=2$ modes of $2.51\pm0.01~\mu {\rm Hz}$, we calculate the rotation period of \astrobj{WD~0246+326} as $3.78\pm0.11$ days.

\subsection{Average Period Spacing}
As shown in Figure 3, the four identified $l=1$ modes are used to make a linear fitting, leading to an average period spacing of $29.3\pm0.2~{\rm s}$.
It is known that the period spacing of pulsating white dwarfs depends sensitively on the total mass of the stars, as shown in Figure 2 of Kawaler \& Bradley (1994)\cite{1994ApJ...427..415K} for pulsating PG1159 stars with models of different masses. Brassard et al. (1992)\cite{1992ApJS...81..747B} also noted this dependence in their Figure 38 for ZZ Ceti stars from a modeling work.

Table 5 lists the average period spacings of the $l=1$, $\Delta P_{l=1}$ modes determined from observations and the stellar mass, $M_*$ deduced from the constraints of the theoretical models for 10 ZZ Ceti stars (Kanaan et al. 2005\cite{2005A&A...432..219K}, Li et al. 2015\cite{2015MNRAS.449.3360L}, Fu et al. 2013\cite{2013MNRAS.429.1585F}, Su et al. 2014\cite{2014MNRAS.437.2566S}, Su, Li \& Fu 2014\cite{2014NewA...33...52S}, Pak{\v s}tien{\.e} et
al. 2013\cite{2013EPJWC..4305012P}, 2014\cite{2014IAUS..301..469P}, C{\'o}rsico et al. 2012\cite{2012MNRAS.424.2792C}, Pech, Vauclair, \& Dolez 2006\cite{2006A&A...446..223P}, Lin, Li \& Su 2015\cite{2015arXiv150105360G}, Bogn\'{a}r et al.\cite{2016MNRAS.461.4059B}). Figure 4 plots the $\Delta P_{l=1}$ versus $M_*$.
A trend of increasing mass with decreasing average period spacing can be seen from Figure 4. With an average period spacing of the $l=1$ modes of 29.3~s, \astrobj{WD~0246+326} would be a massive ZZ Ceti star, with the mass larger than 0.75 solar mass.

\begin{figure}
\resizebox{\hsize}{!}{\includegraphics[angle=90]{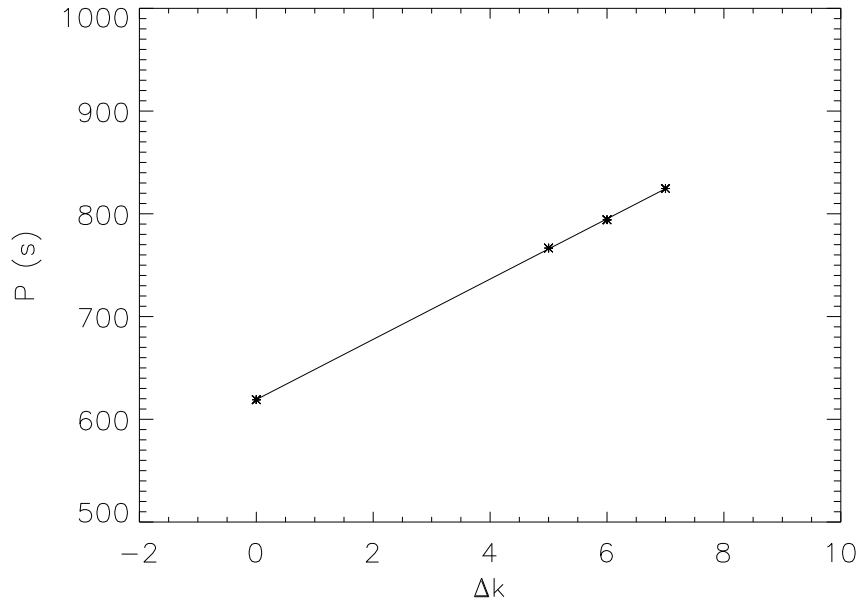}}
\caption{Linear fitting of the four identified $l=1$ modes.}
\label{fig3}
\end{figure}
\section{Constraints from Theoretical Models}
\subsection{Modeling Tool}
We calculated theoretical evolutionary models of white dwarf stars with the WDEC codes (originally developed by Schwarzschild \& H\"{a}rm
1965, see Su et al. 2014\cite{2014MNRAS.437.2566S} for the information of the latest update). Construction of the grid of models is made with the  input of four parameters: the stellar mass, the effective temperature, the Hydrogen mass fraction and the Helium mass fraction.

\begin{table}
\centering
\caption{Summary of average period spacings of the $l=1$ modes determined from observations and the stellar mass deduced theoretical models for 10 ZZ Ceti stars.}
\begin{tabular}{@{}ccc}
\hline ID &  Steller Mass($M_\odot$)&$\Delta P_{l=1}(s)$\\
\hline \astrobj{GD1212}& 0.775&37\\	
\astrobj{KUV~08368+4026} &0.692&49.2\\	
\astrobj{HS~0507+0434B}&0.675&49.63\\	
\astrobj{PG~2303+243}&0.66&	52.0\\	
\astrobj{G117-B15A}&0.593&	55.8\\	
\astrobj{HL Tau 76}&0.575&	49.8\\		
\astrobj{BPM37093}&1.1	&29.4\\	
\astrobj{G207-9}&0.725&	41.7\\
\astrobj{LP~133-144}&0.865&	37\\
\astrobj{KUV~11370+4222}&0.625&58.6\\
\hline
\end{tabular}
\end{table}

\begin{figure}
\resizebox{\hsize}{!}{\includegraphics{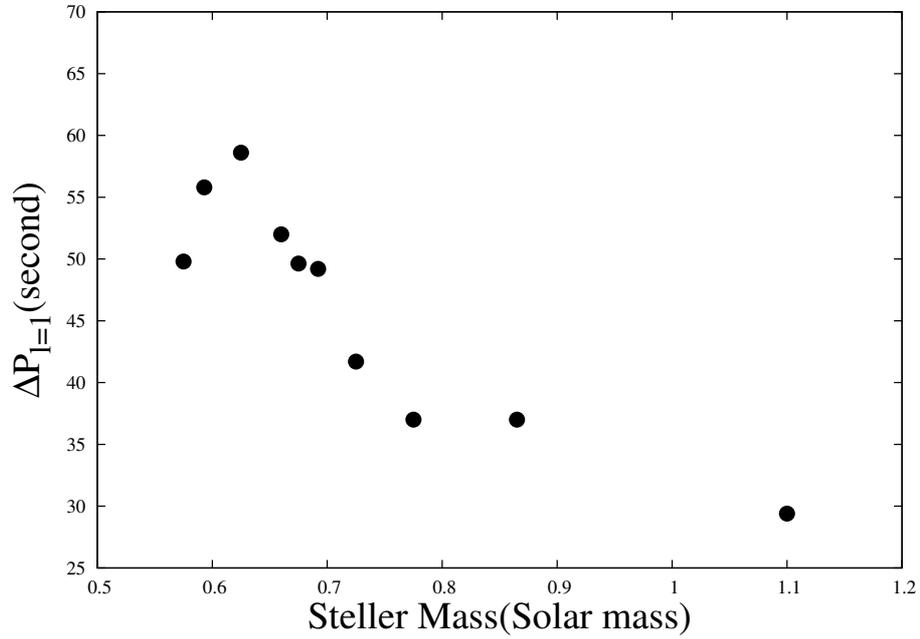}}
\caption{Stellar mass versus average period spacing of l=1 modes for 10 well-studied ZZ Ceti stars.}
\label{fig4}
\end{figure}
\subsection{Stellar Parameters}
Bergeron et al. (1995)\cite{1995ApJ...449..258B} provided for \astrobj{WD~0246+326} the effective temperature $T_{eff}$ of $11290\pm200~{\rm K}$ and the surface gravity  $\log g$ of $8.08\pm0.05$. Gianninas et al. (2011)\cite{2011ApJ...743..138G} presented with spectroscopic observations $T_{\rm eff}=11940\pm180~{\rm K}$ and $\log g = 8.21\pm0.05$. These two sets of parameter are very different to each other and place \astrobj{WD~0246+326} on either the red or the blue edge of the instability strip of ZZ Ceti stars on the H-R diagram.
As indicated in Section 4.2, the value of the average period spacing of the $l=1$ modes determined in present paper suggests that the stellar mass of \astrobj{WD~0246+326} should be larger than 0.75 solar mass. We hence build a coarse grid of models with a wide parameter range and relatively large steps, as listed in Table 6. The Helium mass fraction is fixed as $10^{-2}$ in the grid calculation because the models are not sensitive to this parameter (Brassard et al. 1992)\cite{1992ApJS...81..747B}.

\begin{table}
\centering
\caption{Parameters of the coarse grid of models. Note that $M_H$ is the ratio of the hydrogen layer mass to the stellar mass $M_*$, and $M_{He}$ the ratio of the helium layer mass to $M_*$.}
\begin{tabular}{@{}ccccc}
\hline &\multicolumn{2}{|c|}{Coarse Grid}&\multicolumn{2}{|c}{Fine Grid}\\
\hline  & Range & Step & Range & Step \\
\hline Mass($M_{\odot}$) & $0.75-1.20$  & $0.02$& $0.9-1.02$  & $0.01$\\
$ T_{\rm eff}$($K$) & $11000-14000$ & $200$& $11400-12200$ & $100$\\
 $\log M_{\rm H}$ &$-4.0- -8.0$&$0.5$&$-4.5- -7.0$&$0.1$\\
 $\log M_{\rm He}$ &-2& &-2&\\
 \hline
\end{tabular}
\end{table}

\subsection{Frequency Matching}

Firstly, periods of the four $l=1$ modes, two $l=2$ modes and other six signals with undetermined $l$ (listed in Table 4) are used to match periods of the eigen modes of the theoretical models on the coarse grid. The $\chi^2$ tests are used to select the best fit models as follows,

\[
\chi^2=\sum_n(P^{the}_n-P^{obs}_n)^2
\]
where $P^{the}$ denotes the periods of the eigen modes of the theoretical models and $P^{obs}$ the observationally determined periods.

One minimum of $\chi^2$ was found on the coarse grid, which corresponds to the model of stellar parameters: $M=0.95$~$M_\odot$ and $T_{\rm eff}=11700$~K. In order to put a better constraint on the model parameters, we build a finer model grid with a narrower range of parameters and smaller steps, as listed in Table 6.

Since we can not determined the m values for the frequencies listed in Table 4, we calculate fore each theoretical mode the frequencies of the multiplets due to rotation splitting derived in Section 4.1. Then we search the whole set of frequency for the best fitted ones with the observationally detected frequencies. Frequencies of all identified $l=1$ modes are compared with theoretical ones of the $l=1$modes and their rotation splits, and the same for the $l=2$ modes. The further signals listed in Table 4 are  compared with all $l=1$ and $l=2$ modes and their rotation splits. With the minimum $\chi^2$, one best fit model was found in the procedure. We summarize its stellar parameters in Table 7. Figure 5 displays the distribution of $\chi^2$ test. The three slices are plotted with the hydrogen fraction, the effective temperature and the stellar mass fixed, respectively.

\begin{figure}
\centering
\resizebox{\hsize}{!}{\includegraphics{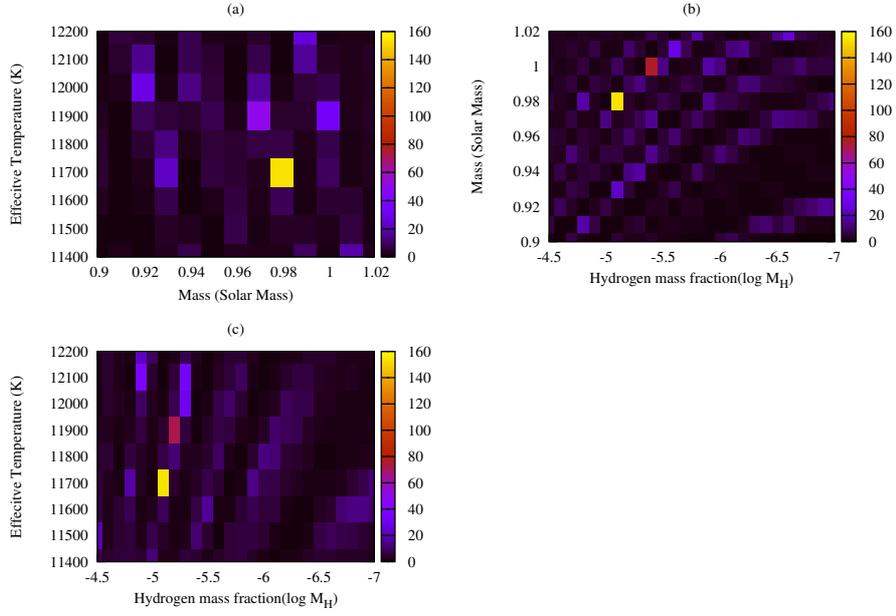}}
\caption{$\chi^2$ test for the fine grid of model. The grey scales on the right side of each panel represent the inversed $\chi^2$ values. (a) Stellar mass versus effective temperature with a fixed hydrogen mass fraction of $10^{-5.1}$, (b) Stellar mass versus hydrogen mass fraction with a fixed effective temperature of 11700~K, (c) hydrogen mass fraction versus effective temperature with a fixed stellar mass of 0.98 solar mass.}
    \label{fig5}
\end{figure}

\begin{table}
\centering
\caption{The best fit model for \astrobj{WD~0246+326}. The symbols are the same as in Table 6.}
\begin{tabular}{@{}cc}
\hline $T_{eff}(K)$ &$11700\pm100$ \\
$\log g$ & $8.66\pm0.02$\\
$M(M_{\odot})$ &$0.98\pm0.01$ \\
$log$ $M_H$ &$-5.1\pm 0.1$ \\
$log$ $M_{He}$&$-2$\\
\hline
\end{tabular}
\end{table}

\begin{table}
\caption{Mode identifications for \astrobj{WD~0246+326}. $P_o$ is the observed period in seconds, $P_c$ the theoretically calculated period of the best fit models as listed in Table 7.}
\centering
\begin{tabular}{@{}cccc}
\hline  &$P_o(s)$ &$P_c(s)$& $(l,k,m)$ \\

\hline \multicolumn{4}{c}{l=1}\\
\hline
a4&   618.8&   619.5&(1,19,-1)\\
b6&   619.3&   620.1&(1,19,0)\\
a8&   766.7&   766.8&(1,24,0)\\
a6&   794.1&   793.5&(1,25,+1)\\
a9&   824.0&   824.5&(1,26,0)\\
a5&   825.3&   825.5&(1,26,1)\\
\hline \multicolumn{4}{c}{l=2}\\
\hline
a11&   846.9&   846.6&(2,47,-2)\\
a10&   847.9&   848.4&(2,47,-1)\\
a2&   851.6&   852.0&(2,47,+1)\\
a3&   883.6&   884.2&(2,49,-1)\\
a1&   890.7&   890.2&(2,49,+2)\\
\hline \multicolumn{4}{c}{further signal}\\
\hline
b1&  1250.3&  1248.5&(2,68,0)\\
b2&   993.2&   993.2&(1,32,-1)\\
a7&   953.0&   954.6&(2,53,-2)\\
b3&   866.2&   865.8&(2,48,-1)\\
b4&   828.7&   828.2&(2,46,-2)\\
b5&   777.6&   777.8&(2,43,0)\\
\hline
\end{tabular}
\end{table}

The best fit model provides a set of $l$, $k$ and $m$ values corresponding to the observationally determined frequencies. Table 8 lists the observed periods and the calculated periods from the best fit model together with the mode identifications.

\section{Conclusions}
Time series photometric observations were made for \astrobj{WD~0246+326} for eight night during a bi-site observation campaign in October 2014. Analysis of the obtained data leads to 11 frequencies. Combined with the results of Bogn\'{a}r et al. (2009)\cite{2009MNRAS.399.1954B}, 17 frequencies, including one triplet, three doublets, and eight single frequencies, are identified as either $l=1$ and $l=2$ modes or further signals. The rotation split of the $l=1$ modes is derived as $1.52\pm0.05~\mu {\rm Hz}$, hence the rotation period is deduced to be $3.78\pm0.11$~days. With the four identified $l=1$ modes, the average period spacing of the $l=1$ modes is determined as 29.3~s, which implies that \astrobj{WD~0246+326} could be a massive ZZ Ceti star. Theoretical modeling for this star is then performed. The theoretically calculated frequencies are compared with the observed ones with the $\chi^2$ test. The stellar parameters of the best fit model are $M=0.98\pm0.01$~$M_\odot$, $T_{\rm eff}=11700\pm 100$~K and the hydrogen mass fraction $log M_{H}/M=-5.1\pm0.1$. The modes of observed frequencies are hence identified.

Although the number of ZZ Ceti stars is the largest in the four known types of pulsating white dwarfs, the samples with stellar parameters determined precisely through asteroseismology are still very limited (less than a dozen). Our result in determining stellar parameters of \astrobj{WD~0246+326} through asteroseismology shows that \astrobj{WD~0246+326} is a massive ZZ Ceti star. With the mass of \astrobj{WD~0246+326} $0.98\pm0.01 M_{\odot}$ would be a newly-discovered ultramassive ZZ Ceti star ever known after \astrobj{BPM37093} with $M\approx 1.1M_{\odot}$, which could experience core crystallization. The debate about the fraction of the crystallized core of \astrobj{BPM37093} depending on the core chemical composition is still open (Kannan et al. 2005\cite{2005A&A...432..219K}, Brassard \& Fontaine 2005\cite{1992ApJS...81..747B}). We also notice that the mass of the DAV \astrobj{GD518} is derived as $1.20\pm 0.03 M_{\odot}$ based upon spectroscopic observations (Hermes et al. 2013\cite{2013ApJ...771L...2H}). However, this mass value is not yet confirmed by means of asteroseismology with existing photometric data. Determination of the stellar mass of \astrobj{WD~0246+326} from this work with $M_* \approx 0.98\pm0.01 M_{\odot}$ provides a new target for study of potential core crystallization of white dwarf stars. Hence, high quality spectroscopic observations and more multi-site photometric campaigns are needed for further study of \astrobj{WD~0246+326} as an interesting ZZ Ceti star.
\section*{Acknowledgements}
CL, JNF and JS acknowledge the support from the Joint Fund of Astronomy of National Natural Science Foundation of China (NSFC) and Chinese Academy of Sciences though grant U1231202 and NSFC grant 11673003, the National Basic Research Program of China 973 Program, the grant 2014CB845700 and 2013CB834900, and the LAMOST FELLOWSHIP supported by Special Funding for Advanced Users, budgeted and administrated by Center for Astronomical Mega-Science, Chinese Academy of Sciences (CAMS). LFM acknowledges the financial support from the UNAM under grant PAPIIT [grant number 105115], and the DGAPA UNAM via project PAPIIT [grant number 106615]. CL thanks Ms. Bo Zhang for her helpful comments.

\section*{References}

\bibliography{mybibfile}

\end{document}